\documentclass[12pt]{article}
\usepackage[polish,english]{babel}
\usepackage[latin1]{inputenc}
\usepackage{times}
\usepackage[T1]{fontenc}
\usepackage{epsfig}

\newcommand{\p}{\partial}

\begin{document}

\title{Scaling, self-similar solutions and shock waves for V-shaped field
potentials }

\author{H. Arod\'z, $\;\;$ P. Klimas $\;$ and $\;$ T. Tyranowski \\$\;\;$ \\  Institute of Physics,
Jagiellonian University, \\ Reymonta 4, 30-059 Cracow, Poland}

\date{$\;$}

\maketitle

\begin{abstract}
We investigate a (1+1)-dimensional nonlinear field theoretic model with the field potential $V(\phi) = | \phi
|.$ It can be obtained as the universal small amplitude limit in a class of models with potentials which are
symmetrically V-shaped at their minima, or as a continuum limit of certain mechanical system with infinite
number of degrees of freedom. We point out that the model has an interesting  scaling symmetry. One consequence
of that is the existence of self-similar solutions of the corresponding field equations. We also find
discontinuous solutions of shock wave type.
\end{abstract}

\vspace*{2cm} \noindent PACS: 03.50.-z, 11.10.Kk \\
\noindent Preprint TPJU - 8/2005

\pagebreak

\section{ Introduction}

Classical (1+1)-dimensional field theoretic models with nonlinear field equations appear in many problems in
physics. Well-known examples are provided by  sine-Gordon and $\phi^4$ models, which have plenty of
applications: from Josephson junctions to macroscopic mechanical systems, see, e. g., \cite{1, 2}. Very rich
contents of these models have been explored in great detail. In the ultralocal limit, i. e., when gradient terms
in the field energies are neglected, both models can be regarded as describing infinite sets of simple uncoupled
one dimensional mechanical systems: pendulums or anharmonic oscillators, respectively \footnote{Let us notice
that the ultralocal limit yields systems with infinite number of degrees of freedom. It should not be confused
with the infinite wave length limit which leads to systems with one degree of freedom: a single pendulum or
anharmonic oscillator.}. In the present paper we consider a model which in that limit describes an infinite set
of coupled balls bouncing vertically from a floor, but of course this is not the reason for investigating it in
detail. The actual motivation is that the model possesses very interesting scaling symmetry. Furthermore, it
universally describes the small amplitude sector in models with non smooth field potentials. In a sense, it is a
counterpart of  free field models which describe the small amplitude sectors in the case of smooth field
potentials.

Specifically, we consider the (1+1)-dimensional, classical field-theoretic model with the field potential
\begin{equation}
V(\phi) = |\phi|,
\end{equation}
and the corresponding field equation
\begin{equation}
\frac{\partial^2 \phi(\xi,\tau)}{\partial \tau^2} - \frac{\partial^2 \phi(\xi,\tau)}{\partial \xi^2} = -
\mbox{sign}(\phi(\xi,\tau)),
\end{equation}
where $\phi$ is a real scalar field. The $\mbox{sign}$ function has the values $\pm 1$ when $ \phi \neq 0,$ and
0 if $ \phi = 0.$ The potential $V$ reaches its absolute minimum at $\phi =0,$ where it is not differentiable -
the left and right derivatives of $V$ have different values:
\[
\left. \frac{d V(\phi)}{d \phi}\right|_{\phi = 0\pm} = \pm 1.
\]
Hence, the potential is symetrically V-shaped at the minimum. This feature distinguishes our model from the
majority of models of condensed matter physics and particle physics. Notice also that Eq.(2) can not be
linearized even if $\phi$ is arbitrarily small.

The non-smooth field potential (1) appears in an effective field theoretic description (in a continuum
approximation) of a discrete mechanical system composed of balls, which are connected by elastic strings and
bounce vertically from the floor, see Fig. 1. Another system with a V-shaped potential - elastically coupled
pendulums bouncing between two rods - is described in \cite{3}. Moreover, V-shaped field potentials appear in
theoretical investigations of pinning phenomenon of extended objects, see, e. g.,  \cite{4}. These facts prove
that eventhough the model might look rather exotic, it is not unrelated to the physical world.

The  model is quite intriguing. First, it is distinguished by the above mentioned universality. Second, apart
from the obious 1+1 dimensional Poincar\`e invariance, it possesses the exact  scaling symmetry of the `on
shell' type, which resembles a little bit the approximate scaling symmetry of Navier-Stokes equations for a
fluid in the turbulent regime. This symmetry is the most profound feature of the model. Because of its presence
we may expect that there exist so called self-similar solutions of the field equation, see, e.g., \cite{5} for
the relevant mathematical concepts. Indeed, we have found such solutions. They have quite interesting structure
both from the physical and mathematical viewpoints. Another very interesting feature of the field equation (2)
is that it admits shock waves: finite discontinuous steps of the field $\phi$ which move with the velocities
$\pm 1$ (`the velocities of light') and leave wakes behind them. One more reason for our interest in that model
is that, to the best of our knowledge, effects of the nonlinearity of the $\mbox{sign}(\phi)$ type in field
equation  have not been thoroughly investigated yet.

The plan of our paper is as follows. In Section 2 two derivations of the model with potential (1) are given: (a)
the continuum approximation to the system of bouncing balls, (b) the small amplitude approximation to a general
symmetrically $V$-shaped potential. Section 3 is devoted to the scaling invariance and to the self-similar
solutions. The shock waves are discussed in Section 4. Section 5 contains a summary and certain remarks.

\section{The derivation of the model}
\subsection{ The continuum limit in the system of elastically coupled, vertically bouncing balls}

Balls of mass $m$ can move along vertical poles which are fastened to the floor  at the points $x_i = a i, \;i =
0, \pm1, \pm2, \ldots,$ lying along a straight line. The balls bounce elastically from the floor. Each of them
is connected with its nearest neighbors by straightlinear segments of elastic strings. The system has one degree
of freedom per ball. It is represented by the elevation $h(x_i,t)$ above the floor of the ball at the point
$x_i$ and the time $t$, see Fig. 1.
\begin{figure}[h]
\begin{center}
\includegraphics[height=200pt,width=330pt]{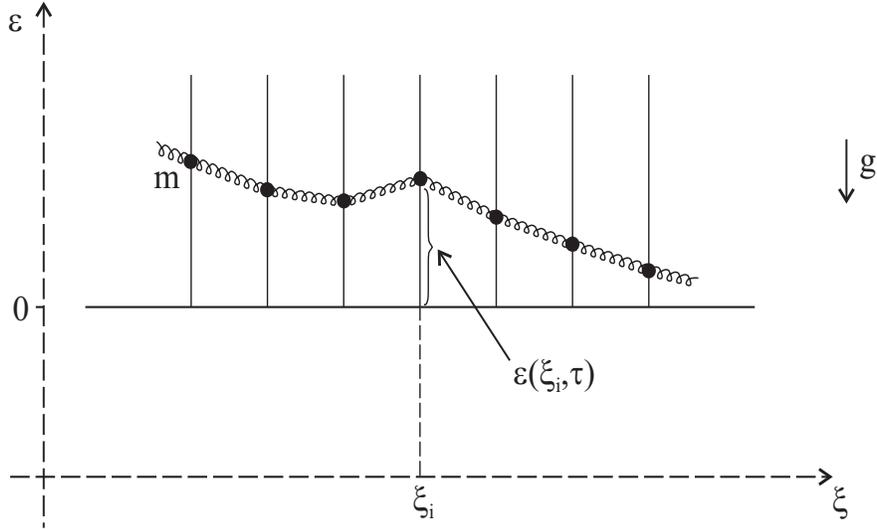}
\end{center}
 \caption{ The  system of balls connected by strings. The balls can move
along the vertical poles (the continuous vertical lines) without friction. They elastically bounce from the
floor which is depicted as the continuous horizontal line. Here $\epsilon \sim h, \;\; \xi \sim x$ and $\tau
\sim t$ are the dimensionless variables introduced in the text. }
 \end{figure}

Let us  forget for the moment about the floor. Then each ball is subject to the force of gravity (represented by
the gravitational acceleration $g$), and to the elastic forces from the two attached strings. It is an
elementary exercise to write the Newton equations of motion for the balls.  Next, we pass to the dimensionless
variables:
\[
\tau = \sqrt{\frac{g}{h_0}} t, \;\; \xi = \sqrt{\frac{mg}{\kappa a^2 h_0}} x, \;\; \epsilon(\xi, \tau) =
h_0^{-1} h(x_i, t), \] where $\kappa$ is the elasticity constant for the strings, $h_0$ is a unit of length
along the poles. Finally,  we take the continuum limit:
\[
a \rightarrow 0, \;\; \frac{m}{\kappa a^2} = \mbox{constans}.
\]
As the result, we obtain the equations of motion for the system of balls in the form
\begin{equation}
\frac{\partial^2 \epsilon(\xi,\tau)}{\partial \tau^2} - \frac{\partial^2\epsilon(\xi,\tau)}{\partial \xi^2} =  -
1,
\end{equation}
provided that $ \epsilon(\xi, \tau) > 0.$ This last condition is due to the fact that we have not taken into
account the presence of the floor.

In case $\epsilon =0$ at certain point $\xi_0$ and time $\tau_0,$   Eq. (2)  is replaced  at that point by the
elastic bouncing condition:
\begin{equation}
\frac{\partial \epsilon}{\partial \tau}(\xi_0, \tau_0) \rightarrow -  \frac{\partial \epsilon}{\partial
\tau}(\xi_0,\tau_0) \;\;\;\; \mbox{when} \;\;\;\; \epsilon(\xi_0,\tau_0) = 0. \end{equation} We can remove this
cumbersome condition with the help of a trick used earlier for the system of pendulums, \cite{3, 6}, and called
 `the unfolding' of the model. In the unfolded model instead of the field $\epsilon(\xi, \tau) \geq 0$ we have
a new field $\underline{\epsilon}(\xi, \tau)$ which can take arbitrary real values. By  assumption, the
evolution equation for $\underline{\epsilon}$ has the form (2),
\begin{equation}
\frac{\partial^2 \underline{\epsilon}(\xi,\tau)}{\partial \tau^2} -
\frac{\partial^2\underline{\epsilon}(\xi,\tau)}{\partial \xi^2} = - \mbox{sign}(\underline{\epsilon}(\xi,\tau)).
\end{equation}
The  field potential corresponding to the r.h.s. of Eq. (5)  has the perfect symmetrically V-shaped form, namely
 \begin{equation}
 \underline{V}(\underline{\epsilon}) = | \underline{\epsilon}|.
 \end{equation}
Because the r.h.s. of Eq. (5) is finite, the velocity $\partial \underline{\epsilon}/\partial \tau$ is a
continuous function of time $\tau$. On the other hand, the two second order derivatives of
$\underline{\epsilon}$ can not be both continuous when $ \underline{\epsilon} =0$ because the r.h.s. of Eq.(5)
is discontinuous.

The original field $\epsilon$ is related to  $\underline{\epsilon}$ by the following formula (the folding
transformation)
 \begin{equation}
 \epsilon(\xi, \tau) = |\underline{\epsilon}(\xi, \tau)|.
 \end{equation}
It is clear that Eq. (5) and formula (7) imply Eq. (3) for $\epsilon$ when $\epsilon \neq 0$. The elastic
bouncing condition (4) follows from formula (7) and from the continuity of $\partial
\underline{\epsilon}/\partial \tau$:
\[
\partial \epsilon/\partial \tau = \mbox{sign}(\underline{\epsilon}) \partial \underline{\epsilon}/\partial
\tau,
\]
hence the sign of $\partial \epsilon/\partial \tau$ changes when $\underline{\epsilon}$ (and $\epsilon$) passes
through 0.

The derivation of Eq.(5) presented above applies also to the phenomenon of pinning of an elastic string (e.g., a
vortex) by a rectilinear impurity. In that case the gravitational force is replaced by the pinning force, and
the system of balls and springs by the string.

\subsection{ The small amplitude sector for the symmetrically V-shaped potential}

 Let us consider a general field equation of  the form
\begin{equation}
\frac{\partial^2 \psi(\xi,\tau)}{\partial \tau^2} - \frac{\partial^2\psi(\xi,\tau)}{\partial \xi^2} + V'(\psi) =
0,
\end{equation}
where the field potential $V(\psi)$ is V-shaped at the minimum located at $\psi = 0,$ see Fig. 2.

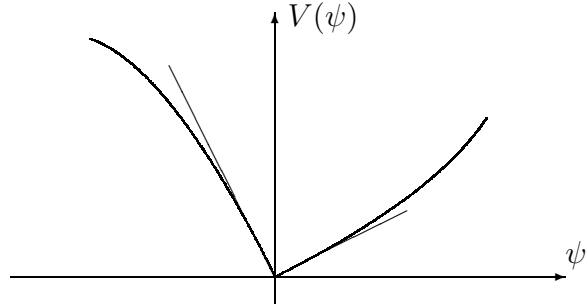
\begin{figure}[tbph!]
\begin{center}
\begin{picture}(200, 110)
\put(0,10){\line(1,0){200}} \put(200,10){\vector(1,0){10}} \put(100,0){\line(0,1){100}}
\put(100,100){\vector(0,1){10}} \put(210,15){\mbox{$\psi$}} \put(105,105){\mbox{$V(\psi)$}}
\put(100,10){\line(2,1){50}} \qbezier(100,10)(160,40)(180,70) \put(100,10){\line(-1,2){40}}
\qbezier(100,10)(60,90)(30,100)
\end{picture}
\end{center}
\caption{Generic V-shaped potential and the piecewise linear
approximation.}
\end{figure}

We are interested in the  small amplitude oscillations of the field $\psi$ around the minimum of the field
potential $V$. Because  $\psi \approx 0,$ we may replace $V'$ by its piecewise linear approximate  form
\[
V(\psi)\cong V'(0+)\: \psi \: \Theta(\psi) - |V'(0-)|\: \psi \: \Theta(-\psi),
\]
where $\Theta$ denotes the step function.  Then, Eq. (8) is replaced by
\begin{equation}
\frac{\partial^2 \psi }{\partial\tau^2} - \frac{\partial^2
\psi}{\partial\xi^2} = - V'(0+) \:\Theta(\psi) + |V'(0-)|\:
\Theta(-\psi).
\end{equation}
Notice that this equation remains nonlinear for arbitrarily small $\psi$ -- there is no linear regime for the
small oscillations! After rescaling the field $\psi$,
\[
\psi(\xi, \tau) = V'(0+) \eta(\xi, \tau),
\]
we obtain the following equation
\begin{equation}
\frac{\partial^2 \eta }{\partial\tau^2} - \frac{\partial^2 \eta}{\partial\xi^2} = - \mbox{sign}(\eta) +
 \left(\frac{|V'(0-)|}{ V'(0+)} -1\right) \:\Theta(-\eta).
\end{equation}
Here we have used the formula $ \Theta(V'(0+) \eta) = \Theta(\eta)$ which is valid because $V'(0+) > 0$. In the
particular case of  symmetrically V-shaped potential $ |V'(0-)| =  V'(0+)$ and we obtain Eq. (2).

Example of physical system for which Eq. (2) can be applied only in the small amplitude limit can be found in
\cite{3}. It is composed of elastically coupled pendulums which bounce from a rod.

\section{The scale invariance and the self-similar solutions}

Equation (2) has the scaling symmetry: if $\phi(\xi, \tau)$ is a solution of it, then
\begin{equation}
\phi_{\lambda}(\xi, \tau) \stackrel{df}{=} \lambda^2 \; \phi(\frac{\xi}{\lambda}, \frac{\tau}{\lambda}),
\end{equation}
where $ \lambda > 0$ is an arbitrary positive number, obeys Eq. (2) too \footnote{Also Eq. (10) has such
symmetry.} . The field energy
\[
E[\phi] = \frac{1}{2} \int d\xi \: [ (\partial_{\tau}\phi)^2 + (\partial_{\xi}\phi)^2 ] + \int d\xi \:V(\phi)
\]
scales as follows:
\[
E[\phi_{\lambda}] = \lambda^3 E[\phi].
\]

When $\lambda \rightarrow 0 $ the  solutions $\epsilon_{\lambda}(\xi, \tau)$ are in general characterized by
high frequencies, short wavelengths and small energies. This is in contrast to, e.g., the massless $\phi^4$
model which also has a scaling symmetry, but in that case the scaling  points to low frequencies and large
wavelengths when the energy $E[\phi_{\lambda}] \rightarrow 0.$ Let us also remark that this reminds the
phenomenon of turbulence -- even more so when we recall that the Navier-Stokes equations in the high Reynolds
number regime have a scale invariance similar to the one described above, see, e. g., the Introduction section
in \cite{7}.

The immediate consequence of the scale invariance is the lack of a characteristic frequency or energy scale. For
example, one can find infinite periodic running wave solutions to Eq. (2) which have the dispersion relation of
the form
\[
\omega^2 - k^2 = \mu^2,
\]
where $\mu^2$ can take any value from the interval $(0, \infty),$ \cite{3}.

Let us stress that the symmetry transformation (11) is defined only in the space of solutions of the field
equation (2).  The action functional corresponding to that equation,
\[
S[\phi] = \int d\tau d\xi \; \left[ \frac{1}{2} (\p_{\tau}\phi)^2 -  \frac{1}{2} (\p_{\xi}\phi)^2 - | \phi |
\right],\] which can be calculated also for fields which do not obey Eq. (2), is not invariant with respect to
the transformations (11). Therefore, the symmetry (11) is not of the Noether type. In such a case one often says
that the symmetry is of the `on shell' kind, see, e. g., \cite{8}.

As always when there is a symmetry, one can find solutions of the pertinent field equation which are invariant
under the symmetry transformation.  In order to find the self-similar solutions of the field equation (2), that
is the ones which are invariant under the scale transformations (11), we adopt the following scale invariant
Ansatz for the solution:
\[
 \phi(\xi,\tau) = \xi^2 S(y), \;\;\; y = \frac{\tau}{\xi}.
\]
In consequence, Eq. (2) is reduced to the ordinary differential equation for the function $S(y)$
\begin{equation}
 (1-y^2)\: S'' +2y\:S' -2\:S = - \mbox{sign}(S).
\end{equation}
This equation has the following polynomial solutions:
\[\mbox{when}
\;\;\; S > 0: \;\; \;\;\;\; S(y) =  - \frac{1}{2} \beta (y^2 +1) +
\frac{\alpha}{2} y + \frac{1}{2},
\]
\[
\mbox{when} \:\;\; S < 0:\;\;\;\;\;\; S(y) =   \frac{1}{2} \beta' (y^2 +1) - \frac{\alpha'}{2} y - \frac{1}{2},
\]
where $\alpha, \beta, \alpha', \beta'$ are arbitrary constants. It has also the trivial solution
\[ S_0(y) = 0. \]
The polynomial solutions are valid on appropriate finite intervals of the $y$-axis which are determined by the
conditions $S > 0$ or $S < 0$, respectively.

Let us try to put together the polynomial solutions in order to cover as large as possible interval of the $y$
axis. The resulting solution has the form depicted in Fig. 3, where $S_k, \:k=1,2, \ldots,$ denote the
polynomial solutions. We assume that $S(0) =0$ because $y=0$ corresponds to the spatial infinity, $ \xi = \pm
\infty$, and we would like to minimize the growth of $\phi(\xi, \tau)$ in that region.

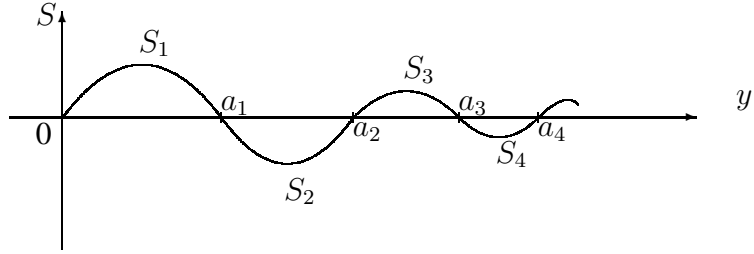
\begin{figure}[tph!]
\begin{center}
\begin{picture}(250, 100)
\put(0,50){\line(1,0){250}} \put(250,50){\vector(1,0){10}} \put(20,0){\line(0,1){80}}
\put(20,80){\vector(0,1){10}} \put(275,55){$y$} \put(10,85){$S$} \put(10,40){0}
 \qbezier(20,50)(50, 90)(80,50)
\qbezier(80,50)(105,15)(130,50) \qbezier(130,50)(150,70)(170,50) \qbezier(170,50)(185,35)(200,50)
\qbezier(200,50)(210,60)(215,55) \put(80,48){\line(0,1){4}} \put(130,48){\line(0,1){4}}
\put(170,48){\line(0,1){4}} \put(200,48){\line(0,1){4}} \put(80,53){$a_1$} \put(130,43){$a_2$}
\put(170,53){$a_3$} \put(200,43){$a_4$} \put(49,75){$S_1$} \put(104,19){$S_2$} \put(149,65){$S_3$}
\put(184,33){$S_4$}
\end{picture}
\end{center}
\caption{The piecewise polynomial solution. $a_k$ are the matching
points for the consecutive polynomials $S_k$.}
\end{figure}

Let us write the polynomial solutions in the form
\[
S_k(y) = \frac{1}{2} (-1)^k [ \beta_k(y^2+1) - \alpha_k y -1].
\]
Equation (12) implies that the following matching conditions at the points $y = a_k$ have to be satisfied:
\begin{equation}
S_k(a_k) = 0 = S_{k+1}(a_k), \;\;\; S'_k(a_k) = S'_{k+1}(a_k),
\end{equation}
provided that $ a_k \neq \pm 1.$ The points $y = \pm 1$ are exceptional because then the coefficient in front of
the second derivative term in Eq. (12) vanishes. This has the consequence that  the first derivative $ dS/ dy $
of the solution does not have to be continuous at the points $y= \pm 1$. Below we shall make use of this
possibility.

The matching conditions (13) give the recurrence relations:
\[
\alpha_{k+1} = \frac{4 a_k}{1 - a_k^2} - \alpha_k, \;\;\;\beta_{k+1} = \frac{2}{1 - a_k^2} - \beta_k,\;\;\;
a_{k+1} = \frac{2 a_k - (1 + a_k^2) a_{k-1}}{1 + a_k^2 - 2 a_k a_{k-1}},
\]
where $\alpha_1 = a_1 >0, \;\; \beta_1= 1, \;\; a_0 =0.$ We also assume that $  a_1 <1$ in order to ensure that
$a_2 > a_1.$ Rather unexpectedly, we have found exact solutions of these relations - it has turned out that a
glance at explicit forms of several initial $\alpha_k, \beta_k$ and $a_k$ is sufficient in order to guess the
solutions. They have the following form
\begin{equation}
\alpha_k = \frac{1}{2} (1+a_1) q^{1-k} \left(1-q^{2k-1}\right), \;\;\; \beta_k = \frac{1}{4} (1 - a_1) \left( 1+
q^{k -1}\right) \left(1+q^{- k}\right),
\end{equation}
\begin{equation} a_k = \frac{1- q^k}{1+q^k},
\end{equation}
where \[  q = \frac{1 - a_1}{1+a_1}.\] Of course, this solution can be checked by substituting it into the
recurrence relations.

Because $ 0 < q < 1,$  formula (15) implies that $ a_k <1,$ and that $  a_k \rightarrow 1 $ when $ k \rightarrow
\infty.$ Therefore, the piecewise polynomial solution constructed above covers only the interval $[0,\:1)$. For
this reason we introduce a special notation for it, namely $S_{pp}(y)$. Simple calculations show that $S_{pp}$
vanishes  when $y \rightarrow 1$. On the other hand, the first derivative $S_{pp}'(y)$ does not vanish in that
limit, but it remains finite.

This solution can be  extended  to the full range of $y$. First, we extend it to the interval $(-1, 1)$ by
taking as the solution  in the interval $(-1,\:0]$ the function $ - S_{pp}(-y),$ which smoothly matches the
function $S_{pp}(y)$ at the point $y=0$. Next, we take the trivial solution $S_0(y) = 0$ for $ y \leq -1$ and $y
\geq 1$. This is possible because $S_{pp}(y)$ and $ - S_{pp}(-y)$ vanish in the limit $ y \rightarrow 1.$ The
first derivatives of the functions $S_{pp}(y)$ and $S_0(y)$  at the points $y= \pm 1$ do not have to be equal
because Eq. (12) allows for a finite jump of the first derivative $S'(y)$ when $y^2 =1.$

The parameter $a_1$ remains free except for the restriction $ 0 < a_1 < 1$, so we actually obtain a family of
self-similar solutions. A snapshot of the solution at certain $\tau > 0$  is depicted in Fig. 4. It is composed
of infinitely many quadratic polynomials in $\xi$ taken on domains which become smaller and smaller when $ \xi
\rightarrow \tau+$ or $ \xi \rightarrow - \tau-$. For  $\xi \rightarrow \pm \infty$  $|\phi|$  approaches the
function $ a_1 \tau |\xi|/2 - \tau^2/2. $  The zeros of $\phi$ lie at $\xi_k(\tau) = \tau / a_k.$ Hence,
$\dot{\xi}_k = 1/a_k >1$ -- the zeros move with `superluminal' velocities. The function $\phi(\xi, \tau)$
exactly  vanishes for $|\xi| < \tau,$ i. e. inside the `light cone'. This
follows from the fact that  the restriction $ -1 < y < 1 $ is equivalent to  $|\xi| > \tau $. \\

\begin{figure}[tbph!]
\begin{center}
\begin{picture}(280,80)
\put(0,30){\line(1,0){250}} \put(125,0){\line(0,1){80}}
\put(250,30){\vector(1,0){5}}\put(125,80){\vector(0,1){5}} \put(129, 80){$|\phi|$} \put(253,20){$\xi$}
\put(85,28){\line(0,1){4}} \put(165,28){\line(0,1){4}} \put(74,20){$\xi = - \tau$}  \put(154,20){$\xi = \tau
>0$} \qbezier(170,30)(175,37)(180,30)
\qbezier(180,30)(190,44)(200,30)
\qbezier(200,30)(220,60)(240,30)\qbezier(240,30)(244,35)(248,40)
\qbezier(168,32)(169,31)(170,30) \put(200,50){\vector(1,0){10}}
\put(85,30){\line(1,0){80}} \qbezier(80,30)(81,31)(82,32)
\qbezier(70,30)(75,37)(80,30) \qbezier(50,30)(60,44)(70,30)
\qbezier(10,30)(30,60)(50,30) \qbezier(2,40)(6,35)(10,30)
\put(60,50){\vector(-1,0){10}}
\end{picture}
\end{center}
\caption{ The self-similar solution. The two arrows indicate that the waves emanate from the points which move
along the $\xi$-axis: $ \xi= \pm \tau$. We plot $|\phi|$ because this is the physical variable in the case of
system of balls - see the relation (7).}
 \end{figure}
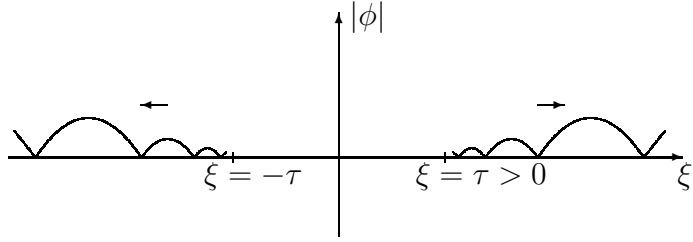

The solution presented in Fig. 4 is symmetric with respect to $\xi \rightarrow - \xi.$ It turns out that the
left- and right-hand halves  of this solution taken separately are solutions too:
 $\phi(\xi, \tau) = \Theta(\xi - \tau)\xi^2 S(\frac{\tau}{\xi}) $ and $
\phi(\xi, \tau) = \Theta(\xi + \tau)\xi^2 S(\frac{\tau}{\xi}) $ also obey Eq. (2).

The full self-similar solution described above can not be physically realized in a physical system. The reason
is that it has infinite energy -- this is the general feature of nontrivial self-similar solutions.
Nevertheless, finite pieces of it can be observed if the dynamics of the model is local, as in our case. Namely,
if a physical finite energy initial configuration differs from our solution only at the very large values of
$|\xi|\;$ $(|\xi| \gg \tau)$, this difference will be seen in the region of finite $|\xi|$ $\;(|\xi| \sim \tau)$
only after certain finite time interval. During that time interval the piece of the self-similar solution
describes the evolution of the finite $|\xi|$ part of the system very well. Information about uses of
self-similar solutions can be found, e. g.,  in \cite{5}.

\section{ The symmetric shock waves}

Another interesting feature of the system governed by Eq. (2) is that it  can support shock waves. Such waves
are not invariant under the scale transformation (11).

 The Ansatz
\begin{equation}
\phi(\xi, \tau) = \Theta(-z) W(z), \;\;\;\; z = \frac{1}{4} (\xi^2 - \tau^2),
\end{equation}
reduces Eq. (2) to the following ordinary differential equation
\begin{equation}
z\:W'' + W' = \mbox{sign}(W),
\end{equation}
where $'$ denotes the derivative $d/dz.$  We have used the following formulas:
\[ \Theta'(-z) = - \delta(z), \;\; z \delta(z) = 0, \;\; z \delta'(z) = - \delta(z), \]
\[
\mbox{sign}(\Theta(-z)
W(z)) = \Theta(-z) \mbox{sign}(W(z)).
\]
Equation (17) has the following partial solutions:
\[\mbox{when} \;\;\; W > 0: \;\;\;\; W_+(z) = z + z_1 + z_1
\:\ln|\frac{z}{z_1}| + d_1,
\]
\[
\;\; \mbox{when} \;\;\; W < 0: \;\;\;\; W_-(z) = - z - z_2 - z_2
 \:\ln|\frac{z}{z_2}| - d_2, \]
where $d_1, d_2, z_1, z_2$ are constants. These solutions are defined on finite intervals of the $z$-axis, see
Fig. 5.

\begin{figure}[tbph!]
\begin{center}
\begin{picture}(150,90)
\put(0,40){\vector(1,0){150}} \put(80,0){\vector(0,1){93}} \qbezier(0,20)(52,80)(75,-10)
\qbezier(90,0)(110,70)(120,80) \put(38,38){\line(0,1){4}} \put(32,30){$-z_1$} \put(38,50){$I$} \put(147,42){$z$}
\put(82,93){$W_+$}
\end{picture}
\hspace*{1cm}\begin{picture}(150,90) \put(0,40){\vector(1,0){150}} \put(80,0){\vector(0,1){93}}
\put(147,42){$z$} \put(82,93){$W_-$} \qbezier(0,60)(52,0)(75,90) \qbezier(90,90)(110,10)(120,0)
\put(32,44){$-z_2$} \put(38,38){\line(0,1){4}} \put(38,24){$II$}
\end{picture}
\end{center}
\caption{The partial solutions $W_{\pm}$. The shock wave is obtained by smooth matching of infinitely many
copies of the parts $I, \:II$. }
\end{figure}
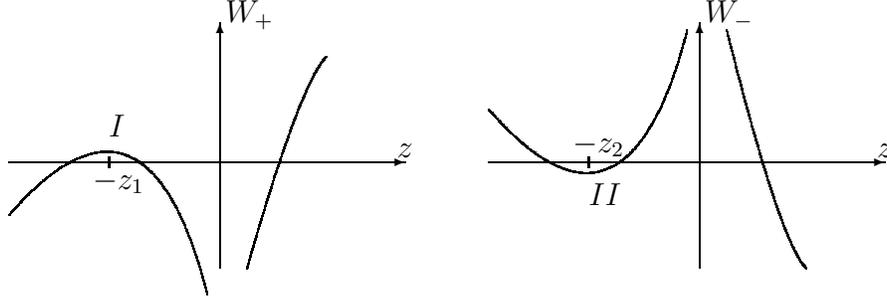

Putting the solutions $W_{\pm}$ together we obtain the solution which is defined for all $z\leq 0$. Because of
the step function in the Ansatz we do not need to know $W(z) $ for $z
>0.$ In order to analyse the matching conditions for the partial solutions we introduce the following notation:
\[ W_{-1}(z) = z + a_0, \;\;\;\; W_k(z) = (-1)^k \left( z + a_k + z_k \ln\frac{|z|}{a_k} \right),
\]
where $ a_k >0.$  The matching conditions at the points $ -a_{k}, \; k = 0, 1, \ldots,$ have the form
\[
(\mbox{a}) \;\;\;\;W_k(-a_k) = 0 = W_{k-1}(- a_{k}), \;\;\; (\mbox{b})\;\;\;\; W'_k(-a_k) = W'_{k-1}(-a_k),
\]
see Fig. 6.

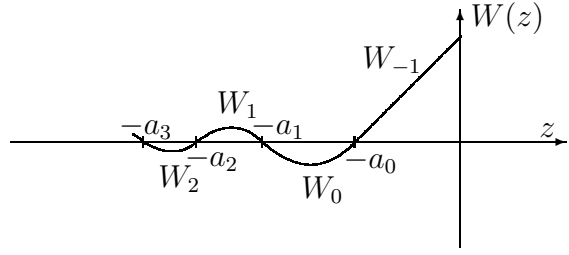
\begin{figure}[tbph!]
\begin{center}
\begin{picture}(240,80)
\put(30,40){\vector(1,0){210}} \put(200,0){\vector(0,1){90}} \put(230,42){$z$} \put(204,84){$W(z)$}
 \put(160,38){\line(0,1){4}}
 \put(125,38){\line(0,1){4}} \qbezier(125,40)(144,23)(160,40)
 \put(100,38){\line(0,1){4}}\qbezier(100,40)(114,51)(125,40)
 \put(80,38){\line(0,1){4}} \qbezier(80,40)(92,33)(100,40)
 \qbezier(160,40)(180,60)(200,80) \qbezier(80,40)(78,42)(76,43)
\put(163,68){$W_{-1}$} \put(140,19){$W_{0}$}
 \put(108,51){$W_{1}$}  \put(86,23){$W_{2}$}
\put(156,30){$-a_0$} \put(121,43){$-a_1$} \put(96,31){$-a_2$} \put(71,43){$-a_3$}
 \end{picture}
 \end{center}
\caption{The scheme for  matching the parts $I,\:II$ of the partial solutions $W_{\pm}$.}
\end{figure}

The matching conditions (b) for the derivatives yield the following recurrence relation
\begin{equation}
z_{k+1} = 2 a_{k+1} - z_k, \;\;\; \mbox{where} \;\;\; z_0 = 2 a_0.
\end{equation}
Here $a_0 > 0$ is a free parameter. The matching conditions (a) for the functions $W_k$ also give a recurrence
relation, but it is much more complicated. In order to write it in a transparent form,  instead of $a_k$ we
equivalently use $x_k$ defined by the following formula
\begin{equation}
a_{k+1} = a_0 \:x_1\: x_2 \cdot \dots \cdot x_{k+1}, \;\;\; x_{k}
>0, \;\; k = 0, 1, \ldots.
\end{equation}
Then, conditions (a) are equivalent to the following infinite set of equations for $x_1, x_2, \ldots$:
\begin{equation}\begin{tabular}{c}
$x_1 = 1 + 2 \ln x_1,$ \\ $x_2 = 1 + 2 \left(1-\frac{1}{x_1}\right) \ln x_2,$ \\  $\vdots$ \\
 $x_k = 1 + 2 \left( 1 - \frac{1}{x_{k-1}} + \frac{1}{x_{k-1}x_{k-2}} - \ldots \pm \frac{1}{x_{k-1}x_{k-2} \cdot
\ldots \cdot x_2 x_1} \right) \ln x_k,$ \\ $\vdots$
\end{tabular}
\end{equation}
These equations can be rewritten in terms of Lambert W function \cite{9}, but we will not give here the related
formulas. Numerical solutions of the first twenty equations (20) are presented in Fig. 7, while Fig. 8 shows the
corresponding values of $a_n$.
\begin{figure}[tbph!]
\begin{center}
\includegraphics[height=5cm, width=8cm]{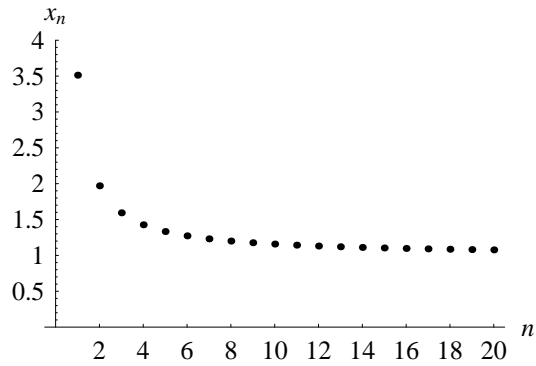}
\end{center}
\caption{ The numerical solutions $x_n$ of Eqs. (20). }
\end{figure}
\vspace*{1cm}

\begin{figure}[tbph!]
\begin{center}
\includegraphics[height=5.5cm, width=8cm]{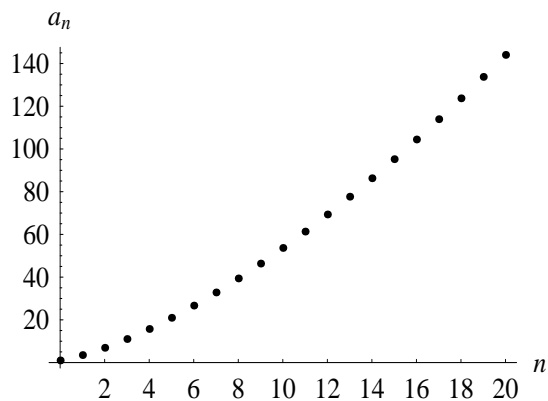}
\end{center}
\caption{ The numerical results for the coefficients $a_n$.}
\end{figure}

The numerical calculations convincingly indicate that $x_n \rightarrow 1$ as $n$ increases. Unfortunately, this
is not sufficient to determine whether $a_n$, related to  $x_k$ by formula (19), increase to  infinity or
converge to a finite value. We have numerically investigated the sums  $ \sum_{n=1}^{N} \ln x_n$ up to $N = 250.
$ The results suggest that the sum becomes divergent when $ N \rightarrow \infty.$ Therefore, we conjecture that
$a_n$ grow to infinity when $n \rightarrow \infty.$ In this case the function $W(z)$ is defined in the whole
half-axis $z \leq 0,$ and formula (16) gives the solution for all values of $z$.

The snapshots of the shock  wave at three times  are shown in Figs. 9, 10, 11.

\begin{figure}[tbph!]
\begin{center}
\begin{picture}(240,60)
\put(-30,15){\vector(1,0){220}} \put(80,0){\vector(0,1){60}} \put(50,15){\line(0,1){30}}
\put(110,15){\line(0,1){30}} \put(45,28){\vector(-1,0){10}}
\qbezier(50,45)(80,10)(110,45)\put(115,28){\vector(1,0){10}} \put(192,5){$\xi$} \put(44,5){$-\tau_0$}
\put(108,5){$\tau_0$} \put(83,57){$|\phi|$}
\end{picture}
\end{center}
\caption{ The symmetric shock wave at an initial time $\tau_0$ such that  $0 < \tau_0 < 2 \sqrt{a_0}$.  The
arrows indicate the directions in which the wave fronts move. $\phi(\xi, \tau_0) =0 $ for $|\xi| > \tau_0.$  }
\end{figure}
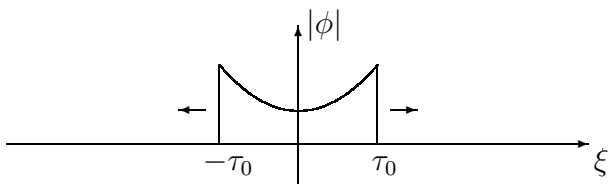

\begin{figure}[tbph!]
\begin{center}
\begin{picture}(240,60)
\put(-30,15){\vector(1,0){220}} \put(80,0){\vector(0,1){60}} \put(20,15){\line(0,1){30}}
\put(140,15){\line(0,1){30}} \qbezier(20,45)(30,25)(65,15) \qbezier(65,15)(80,35)(95,15)
\qbezier(95,15)(130,25)(140,45) \put(193,5){$\xi$} \put(14,5){$-\tau_1$} \put(137,5){$\tau_1$}
\put(83,57){$|\phi|$} \put(15,28){\vector(-1,0){10}} \put(145,28){\vector(1,0){10}}
\end{picture}
\end{center}
\caption{ The symmetric shock wave at a time $\tau_1 > \tau_0$ such that $ 2 \sqrt{a_1} > \tau_1 > 2
\sqrt{a_0}.$ Now $\phi(\xi, \tau_1) =0 $ for $|\xi| > \tau_1.$}
\end{figure}
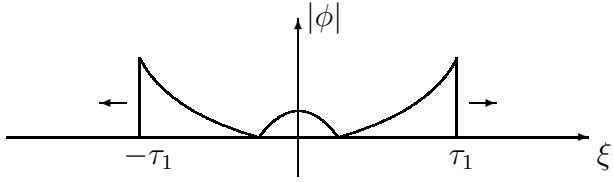

\begin{figure}[tbph!]
\begin{center}
\begin{picture}(240,60)
\put(-30,15){\vector(1,0){220}} \put(80,0){\vector(0,1){60}} \put(10,15){\line(0,1){30}}
\put(150,15){\line(0,1){30}} \qbezier(10,45)(20,25)(45,15) \qbezier(70,15)(80,25)(90,15)
\qbezier(115,15)(140,25)(150,45) \put(193,5){$\xi$} \put(0,5){$-\tau_2$} \put(147,5){$\tau_2$}
\put(83,57){$|\phi|$} \put(0,28){\vector(-1,0){10}} \put(155,28){\vector(1,0){10}} \qbezier(45,15)(57,35)(70,15)
\qbezier(90,15)(102,35)(115,15)
\end{picture}
\end{center}
\caption{ The symmetric shock wave at a time $\tau_2 > \tau_1.$ At later times more zeros of $\phi$ appear.}
\end{figure}
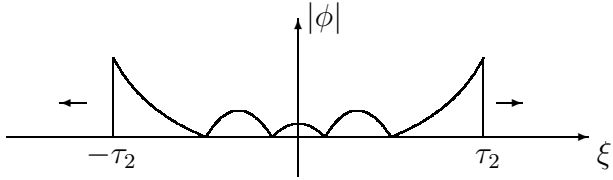

The  features of the shock wave shown on Figs. 9 - 11 can be obtained as follows.    The condition $ z \leq 0$
implies that $\phi(\xi, \tau)$ vanishes outside the `light cone', i. e., when $\xi^2
> \tau^2.$ The 'light cone' corresponds to $z=0$, where $\phi = a_0$. The constant $a_0 >0$ is a free parameter. It gives
the height of jump of the value of the field $\phi$ on the `light cone'. Thus,  $\Theta(-z) W(z)$ describes a
shock wave, symmetric with respect to $\xi \rightarrow - \xi$ and restricted to the `light-cone'. The velocities
of the steps (shock fronts) are equal to  $ \pm 1$.

At the given time $\tau > 0$ the physically visible range of $z$ is finite - it is given by the inequalities  $-
\tau^2/4 \leq z \leq 0.$ Therefore, we can see only a finite number of zeros of $\phi$, but that number grows
indefinitely with time. The positions of the zeros of the field $\phi$   are determined from the equations $ z =
- a_k$. Hence, they are given by the formulas
\[\xi_k(\tau) = \pm \sqrt{\tau^2 - 4 a_k}. \]
Their absolute velocities are greater than 1. Thus, the zeros  move behind the front and faster than it, but
they never catch up with the front. The first zero of $\phi$ appears at the point $\xi =0$ at the time $\tau_i =
2 \sqrt{a_0}$. Notice that it is  by factor $\sqrt{2}$ larger than the time $\sqrt{2a_0}$ needed by a freely
falling ball in order to hit the floor if its initial velocity is equal to zero and the initial elevation above
the floor  to $a_0$. The reason for this difference is obvious: the ball located at $\xi=0$, which indeed has
the elevation $a_0$ and the vanishing velocity in the limit $\tau \rightarrow 0 $, does not fall freely because
it is coupled to its neighbours.

The discontinuous shock waves  have infinite energy because the gradient energy density at the shock fronts is
infinite.  Nevertheless, similarly as the self-similar solutions, they can be very helpful if used with due
care.

\section{ Summary and remarks }

The nonlinear term $  - \mbox{sign}(\phi)$ in the evolution equation (2) rarely is considered in field theory.
Nevertheless, it is quite interesting, in particular because it has the scaling symmetry (11),  and in
consequence  allows for the self-similar waves.  Furthermore, we have shown in an earlier paper, see \cite{3},
that in other models with nonlinearity of that kind there exist static kinks which have strictly finite
extension (no exponential tails) - for that reason they are called compactons.

Systems with non-smooth potentials are quite common in classical mechanics. Especially interesting are so called
impacting systems, for example, a harmonically oscillating particle which can hit a wall when it moves
sufficiently far from its equilibrium position. Recently, such systems have been investigated in connection with
chaotic behaviour and so called grazing bifurcation, see, e. g., \cite{10, 11}. In the present paper we have
considered the field-theoretic counterpart of such systems.

There are many interesting possibilities for future investigations of the models with V-shaped field potentials.
Let us point two of them:\\
1. It is not known whether the models of that kind can be integrable. \\
2. What are properties of quantum version of the (1+1)-dimensional model with the potential $V(\phi) = |\phi|$?
We expect that the scaling symmetry (11) is broken in the quantum theory. What is the resulting mass scale? \\

\section{Acknowledgement}
This work is supported in part by the COSLAB Programme of the European Science Foundation.

\end{document}